\documentclass[twocolumn]{revtex4-1}
\usepackage{amsmath,amssymb,graphicx}
\usepackage{lineno}
\usepackage{amsmath,amsfonts,amssymb}
\usepackage{graphicx}
\usepackage{xcolor}
\usepackage{caption}
\usepackage{subcaption}
\usepackage{float}
\usepackage[colorlinks=true, allcolors=blue]{hyperref}
\begin{document}

\title{\textcolor{black}{Concatenation of Kerr solitary waves in Ceramic YAG: application to coherent Raman imaging}}

\author{N. Bagley}
\affiliation{Southern Methodist University, 6425 Boaz Lane, Dallas, Texas, USA, 75205}

\author{S. Wehbi}
\affiliation{XLIM, Université de Limoges, UMR 7252, 123 Avenue A. Thomas, 87060 Limoges, France}
\affiliation{ALPhANOV, Optics and Lasers Technology Center, Institut d'optique d'Aquitaine, Talence, France}

\author{T. Mansuryan}
\affiliation{XLIM, Université de Limoges, UMR 7252, 123 Avenue A. Thomas, 87060 Limoges, France}

\author{R. Boulesteix}
\affiliation{IRCER, Université de Limoges, UMR 7315, 12 Rue Atlantis, 87068 Limoges, France}

\author{A. Maître}
\affiliation{IRCER, Université de Limoges, UMR 7315, 12 Rue Atlantis, 87068 Limoges, France}

\author{Y. Arosa Lobato}
\affiliation{University of Santiago de Compostela, Praza do Obradoiro S/N, Santiago de Compostela, 15782 Coruña, Spain}

\author{M. Ferraro}
\affiliation{DIET, Sapienza University of Rome Via Eudossiana 18, 00184 Rome, Italy}

\author{F. Mangini}
\affiliation{DIET, Sapienza University of Rome Via Eudossiana 18, 00184 Rome, Italy}

\author{Y. Sun}
\affiliation{DIET, Sapienza University of Rome Via Eudossiana 18, 00184 Rome, Italy}

\author{K. Krupa}
\affiliation{Institute of Physical Chemistry, Polish Academy of Sciences, ul. Kasprzaka 44/52, 01-224 Warsaw, Poland}

\author{B. Wetzel}
\affiliation{XLIM, Université de Limoges, UMR 7252, 123 Avenue A. Thomas, 87060 Limoges, France}

\author{V. Couderc}
\affiliation{XLIM, Université de Limoges, UMR 7252, 123 Avenue A. Thomas, 87060 Limoges, France}

\author{S. Wabnitz}
\affiliation{DIET, Sapienza University of Rome Via Eudossiana 18, 00184 Rome, Italy}

\author{A. Aceves}
\affiliation{Southern Methodist University, 6425 Boaz Lane, Dallas, Texas, USA, 75205}

\author{A. Tonello}
\affiliation{XLIM, Université de Limoges, UMR 7252, 123 Avenue A. Thomas, 87060 Limoges, France}

\email[]{Corresponding author: nbagley@smu.edu} 

\begin{abstract}
A coherent concatenation of multiple solitary waves may lead to a stable infrared and visible broadband filament in ceramic YAG polycrystal. This self-trapped soliton train is leveraged to implement self-referenced multiplex coherent anti-Stokes Raman scattering imaging. Simulations and experiments illustrating the filamentation process and the concatenation of focusing-defocusing cycles in ceramic and crystal YAG are presented. In addition, our simulations and experiments further examine the dependence of the filamentation onset location and supercontinuum generation upon peak input power. Understanding this dependence is key for implementation of viable CARS imaging techniques, \textcolor{black}{due to the comparatively exceptional ability of YAG to generate supercontinuum which can enable higher-sensitivity imaging without delay lines}.
\end{abstract}

\maketitle

\textbf{\textit{Introduction}}
Spatial solitons are formed by the deformation-free propagation of a wave packet, confined by the nonlinear effect of its own intensity \cite{couairon}. Nonlinear saturation allows for wave stabilization over several Fresnel lengths, leading to a modulation-free, axially-symmetric beam. So-called Townes solitons can exactly balance diffraction via nonlinear effects at a given peak power. In two or more spatial dimensions, additional defocusing effects, such as plasma defocusing and nonlinear absorption, are known to arrest “catastrophic collapse” \cite{couairon,silva, replenish}. This process gives, at best, quasi-stability and can otherwise lead to cycles of focusing-defocusing events in which the Townes soliton is never entirely realized; instead, focusing-defocusing cycles of a solitary wave represent non-uniform oscillations around this Townes soliton state \cite{replenish, yanreplenish, ourcleo1, viv1, viv2,  ourcleo2}. It is also known that the extreme confinement of the field produces conical waves, causing some of the beam energy to propagate in a cone shape towards the center and feed the core filament due to a coherent interference effect \cite{couairon, ourcleo1, ourcleo2}. \\ \indent
In this work, we show experimentally and numerically that the self-focusing process obtained in a Nd:YAG ceramic rod can give rise to coherent, concatenated solitary waves, producing a stable filament in both visible and infrared spectra. This stabilization is obtained thanks to the particular dielectric properties of ceramics, which facilitate emission of conical waves, plasma defocusing, and nonlinear/plasma absorption \cite{ourcleo1,aesrome}. Such broadband supercontinuum (SC) is surprisingly stable, and can be efficiently leveraged for self-referenced multiplex coherent anti-Stokes Raman scattering (SR-M-CARS) imaging \cite{cars, ourcleo1}. In particular, for imaging and other broader areas of application, there has recently been considerable interest in extending the length of a filament, or generating a "superfilament" \cite{superfilament}. By enhancing our understanding of the dynamics of focusing-defocusing cycles in YAG and their concatenation, we envisage to design experiments which may produce longer filaments in the laboratory.

\begin{figure}
\centering
\includegraphics[width=0.9\linewidth]{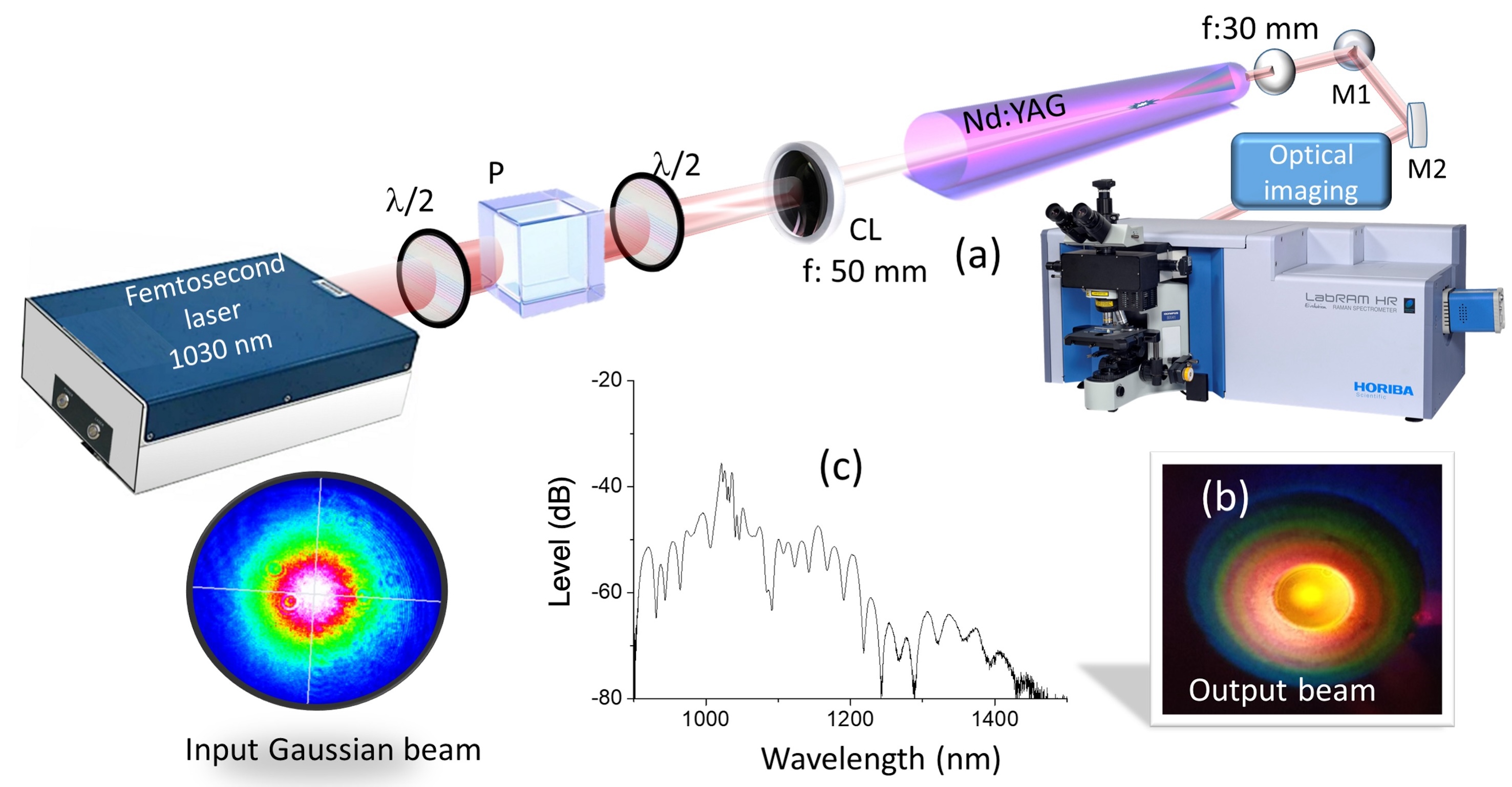}
\caption{(a) Experimental setup composed of a femtosecond laser at 1030 nm, two half waveplates, a polarizer cube, a converging lens and a Nd: Ceramic YAG; (b) output far field \textcolor{black}{of the experiment in} the visible domain; (c) \textcolor{black}{experimental} output infrared spectrum (7 MW \textcolor{black}{peak input power}).
}
\label{fig2}
\end{figure}

\textbf{\textit{Experimental Setup}}
Transparent Nd:YAG ceramics in the form of barrels were elaborated by reaction-sintering according to the process detailed in Ref. \cite{remy}. In our experiments, a 50-mm long Nd:YAG ceramic is excited by a 1030 nm laser source (\textcolor{black}{Pharos (Spark Diadem) emitting 175 (300) fs pulses at 100 (300) kHz}), by using a 50 mm focal lens which creates a 50 $\mu$m diameter Gaussian beam. Two $\lambda$/2 plates and a polarizer cube are placed in front of the crystal, to control the input beam energy and polarization state. At the output, a CCD camera and an optical spectrum analyzer record the shape of the beam and its wavelength content, respectively. The experimental setup is shown in Fig. \ref{fig2}a.
In a second stage, the generated \textcolor{black}{\textbf{infrared}} SC \textcolor{black}{on-axis} is used to perform an SR-M-CARS experiment.
For this application, the beam passes through an LP1000 nm high-pass filter to remove the \textcolor{black}{\textbf{visible}} part of the SC, which would otherwise overlap with the anti-Stokes wave. A slightly tilted 1064 nm notch filter is also used to \textcolor{black}{create a dip in the SC, and reduce} the spectral width of the pump wave, which permits us to obtain sufficient spectral resolution for the SR-M-CARS experiment. The incident beam is then focused on the sample by using a 60$\times$ microscope objective (Olympus LUMFLN60$\times$W), while the SR-M-CARS signal is collected by another 60$\times$ objective (Nikon S Plan Fluor ELWD 60$\times$). This emitted signal is then filtered by a SP1000 nm low pass filter, and analyzed by a HORIBA LabRAM spectrometer. 

 The spatial stability of the solitary wave makes it possible to obtain a sub-$\mu$m \textcolor{black}{transverse} resolution at 1030 nm, while the introduction of a notch filter at 1064 nm guarantees a spectral resolution of less than 40 c$\text{m}^{-1}$.
Fig. \ref{fig2}b shows conical waves and radial self-phase modulation with various interferences, demonstrating the coherence-preserving nature of nonlinear propagation. These interferences are also visible in the infrared spectrum, which is clearly modulated, as shown in Fig. \ref{fig2}c.

\textbf{\textit{Numerical model}}
Propagation of an ultrashort pulse envelope $\tilde{\mathcal{E}}(k_x,k_y,\omega) = \mathcal{F}_{xyt}[\mathcal{E}(x,y,t)]$ in YAG can be described by the following forward envelope equation (FEE) with Kerr nonlinearity, nonlinear absorption, and plasma effects, under the paraxial (small-angle) approximation and the slowly-varying-envelope approximation (in the nonlinear term) \cite{guide, silva, nee1, nee2}
\begin{align}
    \frac{\partial \tilde{\mathcal{E}}}{\partial z} = i \left( \frac{n(\omega) \omega}{c} - \frac{n(\omega_0) \omega_0}{c}  -  \frac{\omega-\omega_0}{v_g}-  \frac{c}{2} \frac{k_x^2 + k_y^2}{n(\omega) \omega} \right)\tilde{\mathcal{E}} \nonumber \\ + i \frac{\omega_0 }{c} n_2 \mathcal{F}_{xyt}[\mathcal{I} \mathcal{E}]  - \frac{\sigma(\omega)}{2} \mathcal{F}_{xyt}[\rho  \mathcal{E}]  - \frac{\beta_{17}}{2} \mathcal{F}_{xyt}[\ \mathcal{I}^{16}   \mathcal{E}]
    \label{eqn:eqn1}
\end{align}
Here, $v_g$ is the group velocity evaluated at $\omega = \omega_{0}$ and $\beta_{17} = 17 \omega_0 \sigma_{17} \rho_{nt}$ and  $\mathcal{I} = (\epsilon_0 c n_0/2) |\mathcal{E}|^2 $ and $\sigma(\omega) = q_e^2 \tau_c[n(\omega) c \epsilon_0 m_e]^{-1} (1 + i \omega \tau_c)[1 + \omega^2 \tau_c^2]^{-1}$. We have $\lambda_0 = $ 1030 nm and we select $\lambda_{\text{ref}} = $ 780 nm. We select $n_{2} = 6.13 \times 10^{-20}$ $\text{m}^2$/W. We select $\tau_c$ = $3 \times 10^{-15}$ s and  $\sigma_{17} = 1 \times 10^{-274}$ $\text{kg}^4$ s  $\times \text{Pa}^{-14} \times \text{W}^{-7}$ and $\rho_{nt} = 7 \times 10^{28}$ $\text{m}^{-3}$ and $U_i = 1.04 \times 10^{-18}$ J as in Ref. \cite{silva}.

The split-step method with a step size of $\Delta z = 4$ $\mu$m is used to solve Eq. \ref{eqn:eqn1} \textcolor{black}{using MATLAB}. In the splitting of operators, we first propagate the linear terms in Fourier space. Then, the operator part  corresponding to the second (Kerr) and fourth (nonlinear absorption) terms are applied in real space, and finally the third (plasma) term is applied in real space by integration of this term forward by a distance $\Delta z$ (using the forward Euler method). The rate equation that is solved for \textcolor{black}{the plasma current density,} $\rho$, can be found in, for instance, section 2.4.4 of Ref. \cite{guide}. 

\begin{figure}[H]
\centering
\includegraphics[width=0.9\linewidth]{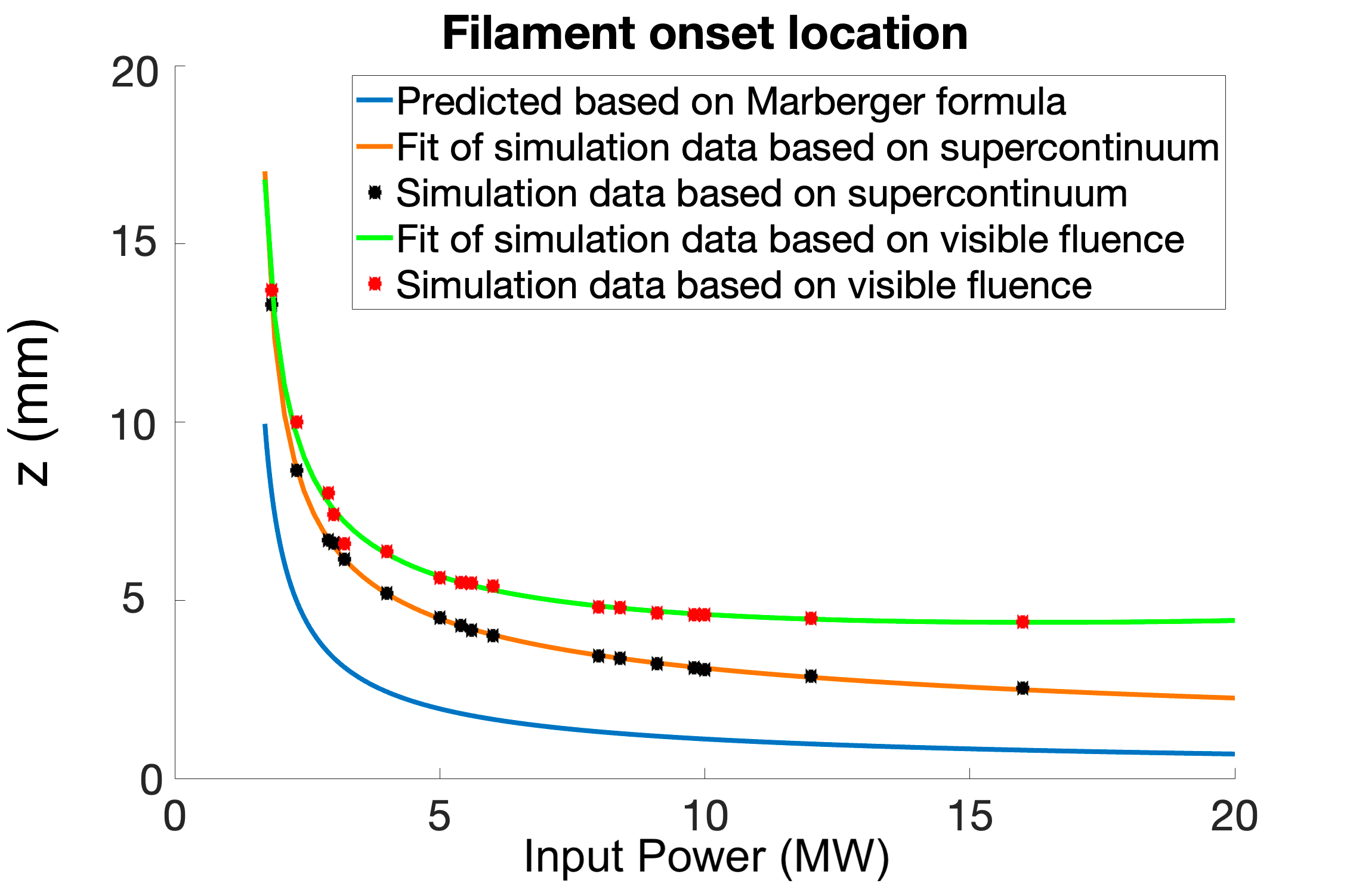}
\caption{Simulated filament onset location (with defocusing terms and 300 fs pulse duration) vs. (Blue): predicted (via Marburger formula, without defocusing). (Green, red): Location where on-axis visible fluence first surpasses .25 J/$\text{m}^2$. (Orange, black): Location where spectral flux density of the rod first surpasses $0.001 \text{W}/\text{Hz}$ at every visible frequency in the 525-750 nm range. }
\label{figfilons}
\end{figure}

\textbf{\textit{Defocusing effect}} 
In the absence of defocusing effects, the critical distance at which catastrophic collapse occurs for a monochromatic pulse is described by the Marburger formula \cite{guide}: $z_{\text{cr}} = z_R[{-.0219 + (-0.852- \sqrt{P_{\text{in}}/{P_{\text{cr}}}})^2}]^{-1/2}$.
In our case, the Rayleigh length is $z_R = 0.005476$ m, and the critical power $P_{\text{cr}}$ is 1.4 MW. Equivalently, the critical distance can be expressed as

\begin{equation}z_{\text{cr}} = [{a_1 P_{\text{cr}} +a_2 P_{\text{cr}}^{1/2} + a_3}]^{-1/2} \label{eqn:predfilons}\end{equation}
with $a_1 = 355.734$, $a_2 = - 716.94 $, and $a_3 = 350.331$. We performed a fit of our spectrally-resolved (with 300 fs pulse duration) simulation data, including the arrest of collapse due to defocusing and spectral broadening. This is shown in \textcolor{black}{Fig.} \ref{figfilons}, where the orange curve, for instance, corresponds to \textcolor{black}{Eq.}\ref{eqn:predfilons} with $(a_1,a_2,a_3)$ = (-4334.23, 47688.41, -51413.18).

\begin{figure}[H]
\centering
\includegraphics[width=0.9\linewidth]{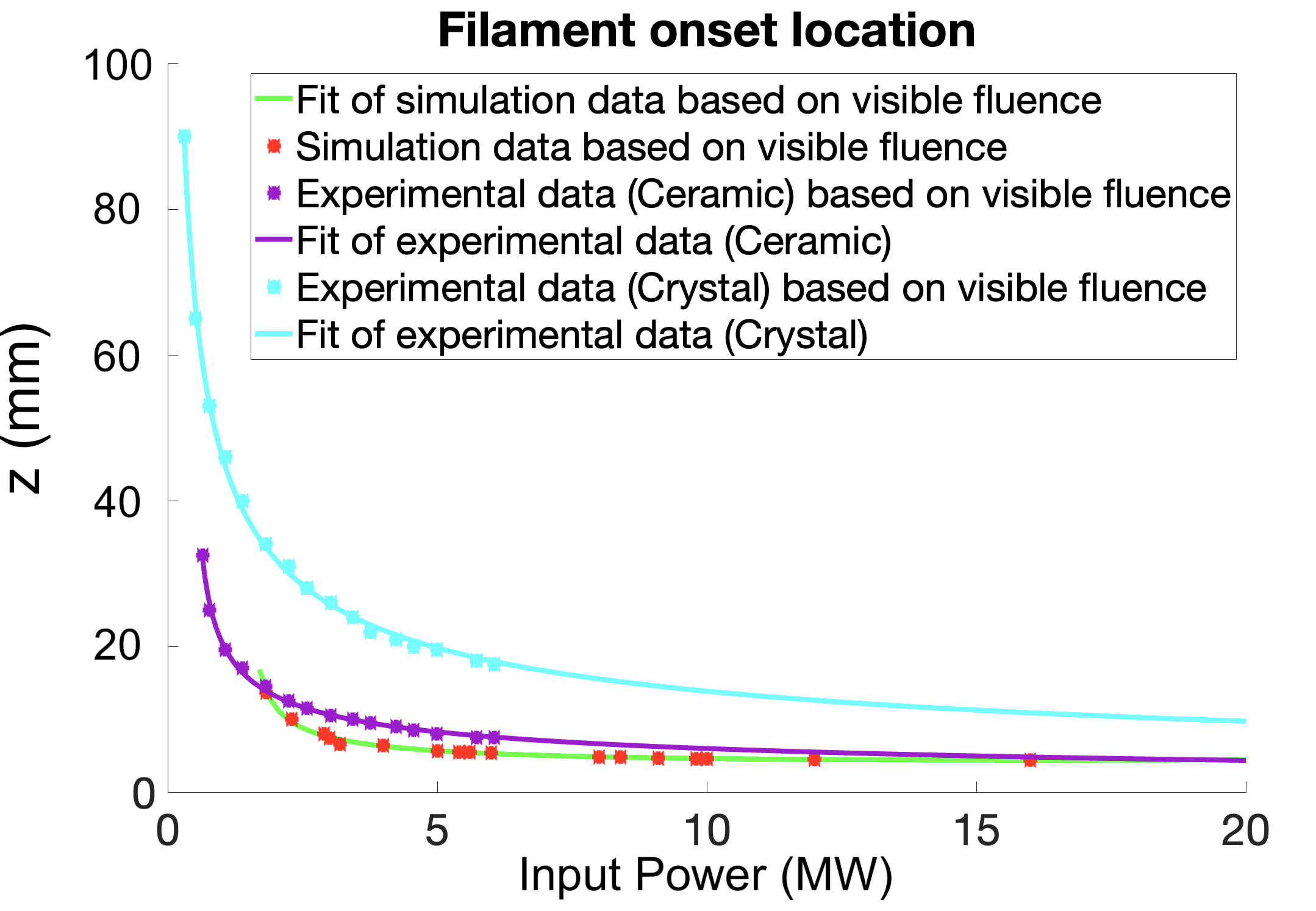}
\caption{300 fs pulse duration: (Green, red) Simulated filament onset location from Fig. \ref{figfilons}. (Cyan): Crystal-YAG experimental data. (Purple): Ceramic-YAG experimental data.}
\label{figfilons2}
\end{figure}
\textcolor{black}{Since plasma effects and supercontinuum generation (SCG) can arrest collapse} \cite{couairon}, the defocusing action (and inclusion of spectrally-resolved effects) shifts the filament onset location further away from the origin than predicted by the Marburger formula. However, as shown in Figs. \ref{figfilons} and \ref{figfilons2}, both experimental and simulated data can be fit with good accuracy by Eq. \ref{eqn:predfilons}, albeit for different values of $a_1, a_2, a_3$. The discrepancy between experiment and simulation (see Fig. \ref{figfilons2}) may be ascribed to the fact that light is \textcolor{black}{focused by a lens} when it first enters the rod in the experiment. 

\begin{figure}
\centering
\includegraphics[width=\linewidth]{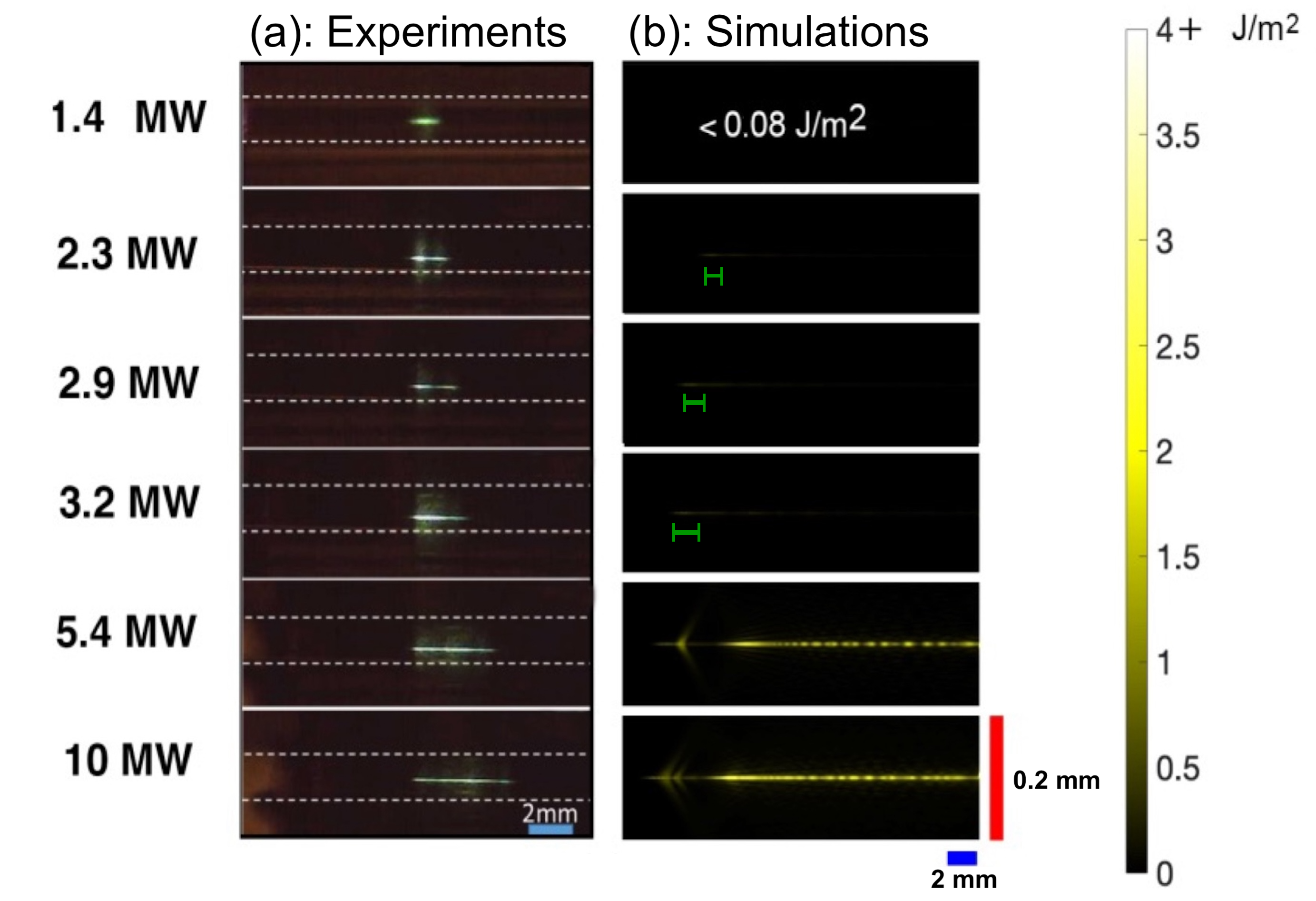}
\caption{Coherent Kerr-soliton beam generated for 175 fs pulse duration: \textcolor{black}{(a) Experimental demonstration of beam concatenation in Nd:YAG ceramics. All filament onset locations have been aligned, to show the progressive lengthening of the solitary wave. These results are comparable to the experimental results for filament onset shown in Fig. \ref{figfilons2} (purple).
(b): \textcolor{black}{Simulated Intensity $\mathcal{I}$, band-pass filtered to include only 380-750 nm and then integrated in time. We use Eq. \ref{eqn:eqn1} to propagate $\mathcal{E}$ along the z-direction.} For visibility, the green brackets highlight the initial focusing cycle in low peak input power cases.} 
}
\label{fig1}
\end{figure}

\begin{figure}
\centering
\includegraphics[width=0.9\linewidth]{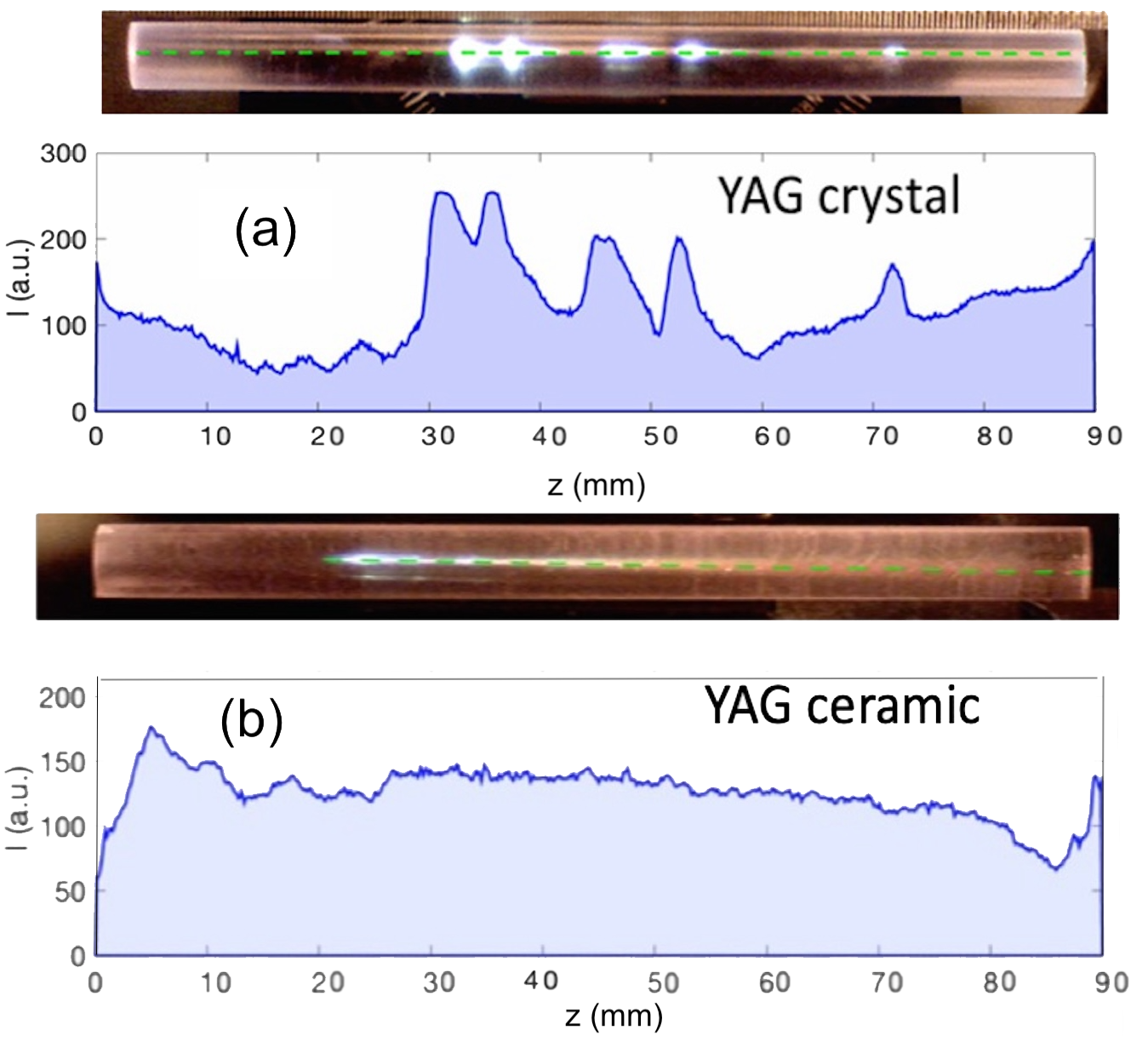}
\caption{Experimental comparison between (a) Nd:YAG crystal and (b) Nd:YAG ceramics at 10 MW peak power for 300 fs input pulse duration. \textcolor{black}{This primarily illustrates the effect of different values of electrical conductivity, which is proportional to the collision time $\tau_c$. For a crystal (panel (a)) we have $\tau_c = 3 \times 10^{-15}$ s \cite{silva}}.} 
\label{fig1newnewnew}
\end{figure}

The capability to control the focal point (and producing a focus that nearly-instantaneously varies with peak input power) is critical for \textcolor{black}{SR-M-CARS imaging \cite{cars}. In fact, the peak input power influences both the SCG and the filament onset location in the ceramics, which in turn affects the focal point of the microscope on the sample. However, the low level of laser fluctuations (less than 4\% in intensity) permits a good shot-to-shot stability of the measurements. Moreover, we expect that the use of YAG ceramics can mitigate the plasma defocusing action due to the different dielectric properties \textcolor{black}{(primarily the electron collision time $\tau_c$) when compared to YAG crystals} \cite{YAGceramicsYu2016, aesrome}.} To study this effect and to supplement Figs. \ref{fig1} and \ref{fig1newnewnew}, we included a video of our experiments in the 300 fs pulse duration case (see Visualizations 1 and 2 of supplementary material), showing the effects of varying the peak input power. 

\begin{figure}
\centering
\includegraphics[width=\linewidth]{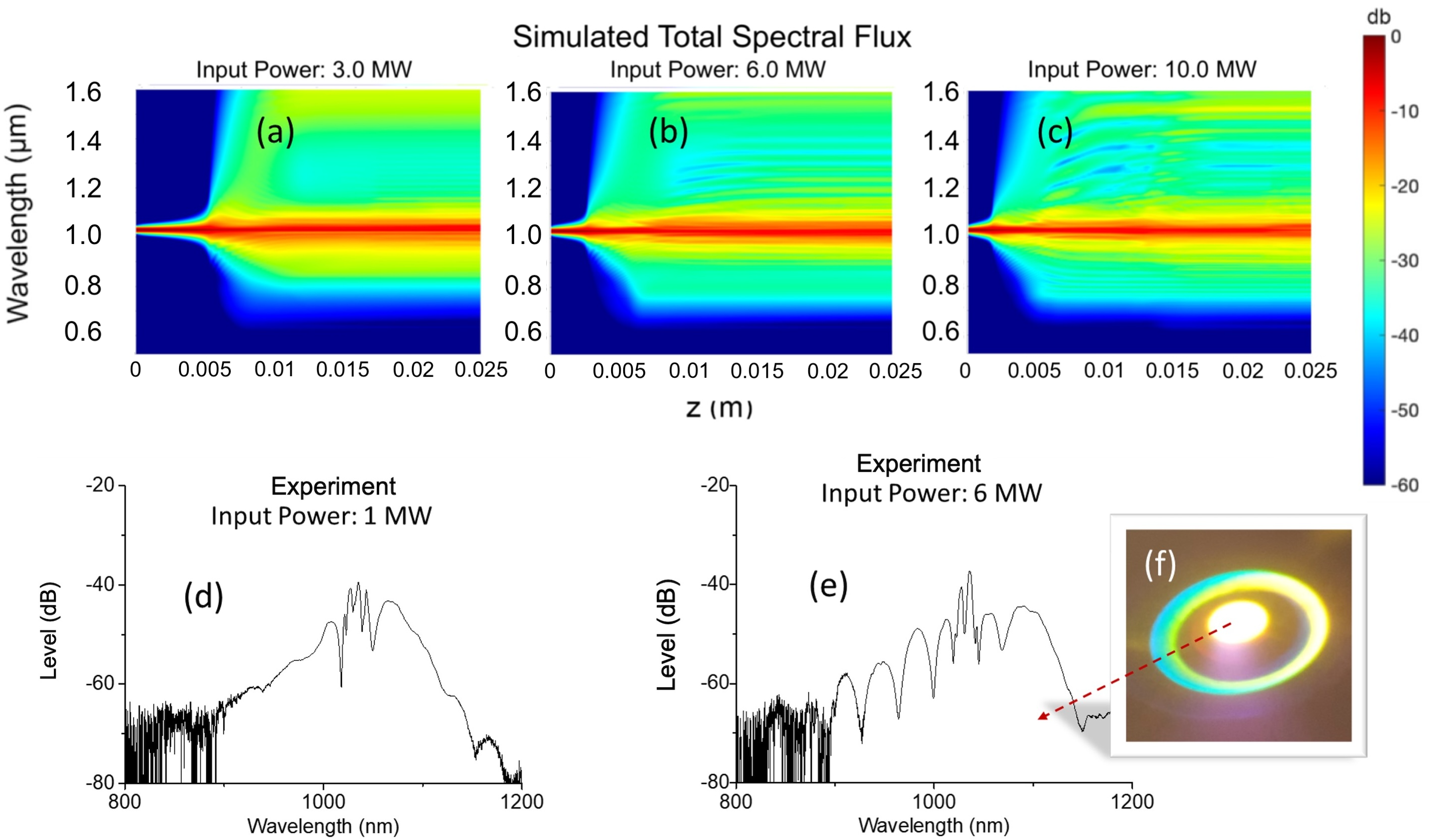}
\caption{\textcolor{black}{Panels (a),(b),(c): numerically simulated SCG at 175 fs and three different peak input powers. The spectral broadening originates at the point of beam collapse where the maximum local intensity is reached. Experimental spectra at the output beam center at 1 MW (d) and 6 MW (e). Panel (f): sample case of output beam shape (far field) at 10 MW (visible spectrum). } }
\label{figsuper}
\end{figure}

\textbf{\textit{Soliton Concatenation}} 
For a peak input power of 1.4 MW (equal to the critical power $P_{\text{cr}}$), a single self-focusing process appears, and spectral broadening accompanies the trapped spatial beam (Fig. \ref{fig1}a). After a few hundred microns of propagation the solitary wave spreads out, owing to the defocusing action of plasma generation and the emission of conical waves. By increasing the peak input power, a second self-focusing process appears, as shown in Fig. \ref{fig1}a. With a further peak input power increase, we observe that the second beam gradually approaches and concatenates with the first, thus substantially increasing the overall propagation length of the initially trapped wave (Fig.\ref{fig1}a). \textcolor{black}{For higher peak powers, the length of the solitary wave exceeds 5 mm, which is 5 times the filament length at 1.4 MW.} 
The concatenation of Kerr-solitons and SCG necessary for SR-M-CARS is also represented by the numerical simulation results shown in Fig. \ref{fig1}b and Fig. \ref{figsuper}a-c, respectively. Notably, a similar experiment carried out with a YAG crystal instead of Nd:YAG ceramics led to different dynamics (see Fig.\ref{fig1newnewnew} and Visualizations 1 and 2 of supplementary material), which keep the individual trapped beams spatially separated .
This suggests that a specially-conceived ceramic YAG with different dielectric properties could generate a \textcolor{black}{more uniform and longer train of solitary waves with enhanced robustness to the peak pump power.}
\textcolor{black}{Plasma rate equation parameters such as the collision time $\tau_c$ (which is proportional to the electrical conductivity) influence the efficiency of plasma generation, particularly the magnitude of the plasma defocusing term in Eqn. \ref{eqn:eqn1}, the maximum intensity (and minimum beam width), and the distance between focusing-defocusing cycles \cite{couairon}. Therefore, the material parameter $\tau_c$ has an important effect on the maximum length of filaments.} 

Additionally, Figure \ref{fig1} shows the formation of a very cylindrically-symmetric beam, which does not drift away from the origin nor produce multiple filaments. This is also the case in experiments, indicating that any inhomogeneities present in the YAG crystal or ceramics do not significantly impact the beam quality. Since the system preserves cylindrical symmetry, a single clean beam is produced at a robust transverse location, which is helpful for \textcolor{black}{SR-M-CARS} \cite{cars}.

\begin{figure}
\centering
\includegraphics[width=\linewidth]{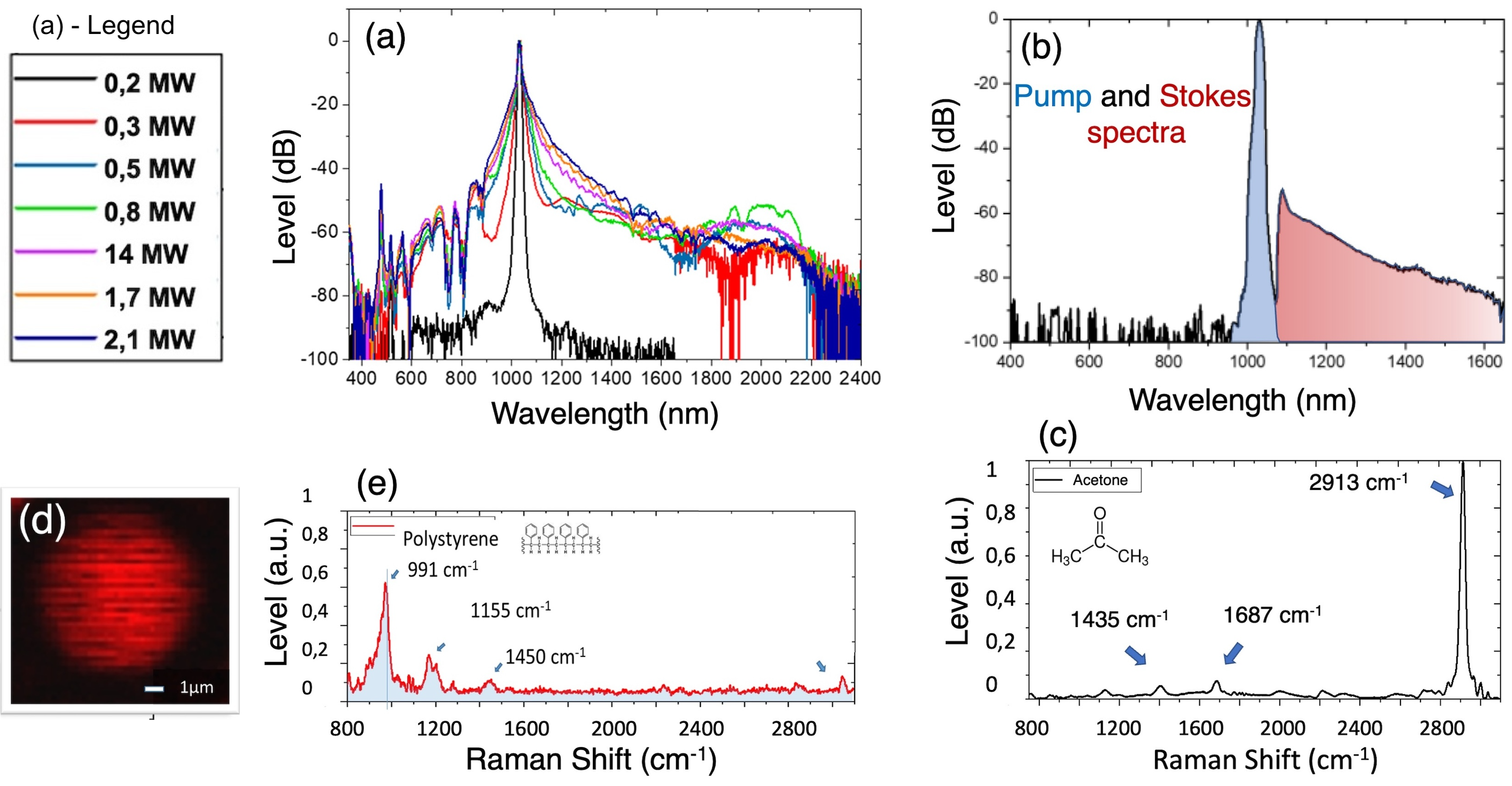}
\caption{\textcolor{black}{175 fs Input pulse duration: (a) SC spectra shown as a function of the pump peak power; (b) pump and Stokes spectra; (c) M-CARS spectrum of Acetone; (d) image of polystyrene bead at 991 $\text{cm}^{-1}$ (e) M-CARS spectrum of polystyrene}}
\label{figCARS}
\end{figure}

\textbf{\textit{Supercontinuum Generation}}
Efficient SCG is essential for the SR-M-CARS process \cite{cars, cars2, cars3}.
\textcolor{black}{Figure \ref{figsuper} shows how controlling wave collapse in Nd:YAG ceramics can lead to efficient coherent SC in the visible and infrared domain. Such nonlinear conversion can be either collinear (emitted by the central part of the beam) or non-collinear (conical wave emission). The spectrum spanning from 1000 nm to 1700 nm allows for CARS multiplexing \cite{cars} between 0 and 3000 $\text{cm}^{-1}$, as shown in Fig. \ref{figCARS}. The SC generated in the Nd:YAG ceramics (see fig.\ref{figCARS}a) is dominated by self-phase-modulation; it covers both the visible and near-infrared regions and decays nearly exponentially on either side of the pump wavelength. }

Numerical simulations (most notably \ref{figsuper}c) show some spectral ripples between 1100 nm and 1600 nm, which is likely due to the spectral interference among the notches. These results are consistent with experimental results (see Fig.\ref{figsuper}\textcolor{black}{d},e). The principal mechanism for this phenomenon is self-phase modulation, illustrating the coherent nature of the trapped waves. These interferences appear in experiments as colored rings around the beam center (see Fig.\ref{fig2}b). \textcolor{black}{While conical emission primarily occurs at red wavelengths (see Fig.\ref{fig2}b), in both experiments and simulations other visible wavelengths exhibit conical emission at varying angles (see Fig.\ref{figsuper}f and Fig. \ref{fig1}b).}   
\\
\indent\textbf{\textit{Conical Emission}}
\textcolor{black}{Such an effect} is well known to occur in cases of filamentation, and it plays an important role in smoothing wave collapse \cite{couairon,mythesis}.  
For instance, simulations in Fig. \ref{fig1}b show the importance of reflected conical emission in feeding the central filament, most obviously in the 5.4 MW and 10 MW peak input power cases. 

\textbf{\textit{Demonstration of CARS}}
By using two filters (LP1000 and a notch filter at 1064 nm ), we were able to isolate a sharp spectral slice at 1030 nm alongside with a broadband Stokes wave, which allows for the generation of a large band SR-M-CARS signal (see Fig. \ref{figCARS}b). The first experiment employed an Acetone sample. We obtained its signature in the form of an asymmetric C$\text{H}_3$ stretching mode at 2913 $\text{cm}^{-1}$, the C=O stretching mode at 1687 $\text{cm}^{-1}$, and the deformation of C$\text{H}_3$ bonds at 1435 $\text{cm}^{-1}$, respectively (see Fig. \ref{figCARS}c). 

By using a sample of polystyrene beads, we proved that a transverse scan can also provide an image with \textcolor{black}{selective M-CARS resonance at 991 c$\text{m}^{-1}$ (see Fig.\ref{figCARS}d).} We obtained the C$\text{H}_2$ scissoring mode at 1450 $\text{cm}^{-1}$, the C-C stretching mode at 1115 $\text{cm}^{-1}$, the ring breathing mode at 991 $\text{cm}^{-1}$, and the aromatic C-H stretching vibration at 3054 $\text{cm}^{-1}$ (see Fig. \ref{figCARS}e), respectively. Further demonstrations of this method \textcolor{black}{and its applications} will appear in a separate publication; some details appear in Re. \cite{wehbithesis} and in Section 2 of Ref. \cite{mythesis}..
\\\\\indent\textbf{\textit{Conclusion}}
\textcolor{black}{In conclusion, we experimentally demonstrated that spatial solitons, generated at different positions in a ceramic polycrystal under the combined action of Kerr self-focusing and plasma defocusing, may merge and produce a continuous and stable laser filament, which propagates over several millimeters. We have presented a study of the defocusing action, filament onset location, SCG and conical emission in crystal and ceramic YAG. 
Moreover, we exploited the broadband output spectrum of the trapped beam with a clean spatial shape for SR-M-CARS imaging experiments. Optimizing the dielectric characteristics of ceramics may additionally make it possible to control the plasma defocusing process, and mitigate spatial instabilities of two-dimensional Kerr solitons}. 
\\\\\\\indent\textcolor{black}{\textbf{\textit{Funding}} Authors acknowledge: French National Research Agency: Labex 10-LABX-0074-01 Sigmalim,
EUR 18-EURE-0017 TACTIC, Equipex  ANR-21-ESRE-0012 ADD4P, ANR-23-CE08-0021-02 3Dmoc and ANR-24-CE46-7295-01 OSMOSIS.
K. Krupa: Narodowa Agencja
Wymiany Akademickiej
(BPN/BFR/2021/1/00013); Campus France
(48161TH).
Y. Arosa Lobato: postdoctoral fellowship (ED481B-2021-027) from the Xunta de Galicia. N. Bagley: NSF GRFP Grant No. DGE-2034834. B. Wetzel: ERC - EU Horizon 2020, GA 950618 - STREAMLINE and Conseil Regional Nouvelle-Aquitaine - SPINAL.
}
\\\textcolor{black}{\textbf{\textit{Disclosures}} The authors declare no conflicts of interest} 

\textcolor{black}{\textbf{\textit{Data availability}} Data underlying the results presented in this paper are not publicly available at this time but may be obtained from the authors upon reasonable request} 

\bibliography{sample}

\end{document}